\documentclass[
aps,prl,
reprint,
a4paper,
superscriptaddress,
longbibliography,
preprintnumbers
]{revtex4-1}
\usepackage[utf8]{inputenc}
\usepackage[T1]{fontenc}

\usepackage{amsmath}
\usepackage{braket}
\usepackage{gensymb}
\usepackage{graphicx}
\usepackage{hyperref}
\usepackage{units}

\hypersetup{citecolor=magenta}
\hypersetup{colorlinks=true}
\hypersetup{linkcolor=blue}
\hypersetup{urlcolor=blue}

\newcommand{\ketbra}[2]{{\ket{#1}\!\bra{#2}}}

\begin{document}

\title{Device-independent certification of a nonprojective qubit 
measurement}

\author{Esteban~S.~Gómez}
\author{Santiago~Gómez}
\author{Pablo~González}
\author{Gustavo~Cañas}
\author{Johanna~F.~Barra}
\author{Aldo~Delgado}
\affiliation{Departamento de Física, Universidad de Concepción, 160-C 
Concepción, Chile}
\affiliation{Center for Optics and Photonics, Universidad de Concepción, 160-C 
Concepción, Chile}
\affiliation{MSI-Nucleus for Advanced Optics, Universidad de Concepción, 160-C 
Concepción, Chile}

\author{Guilherme~B.~Xavier}
\affiliation{Center for Optics and Photonics, Universidad de Concepción, 160-C 
Concepción, Chile}
\affiliation{MSI-Nucleus for Advanced Optics, Universidad de Concepción, 160-C 
Concepción, Chile}
\affiliation{Departamento de Ingeniería Eléctrica, Universidad de Concepción, 
160-C Concepción, Chile}

\author{Adán~Cabello}
\affiliation{Departamento de Física Aplicada II, Universidad de Sevilla, 
E-41012 Sevilla, Spain}

\author{Matthias~Kleinmann}
\affiliation{Department of Theoretical Physics, University of the Basque 
Country UPV/EHU, P.O.~Box 644, E-48080 Bilbao, Spain}

\author{Tamás~Vértesi}
\affiliation{Institute for Nuclear Research, Hungarian Academy of Sciences, 
H-4001 Debrecen, P.O.~Box 51, Hungary}

\author{Gustavo~Lima}
\email{glima{@}udec.cl}
\affiliation{Departamento de Física, Universidad de Concepción, 160-C 
Concepción, Chile}
\affiliation{Center for Optics and Photonics, Universidad de Concepción, 160-C 
Concepción, Chile}
\affiliation{MSI-Nucleus for Advanced Optics, Universidad de Concepción, 160-C 
Concepción, Chile}


\begin{abstract}
Quantum measurements on a two-level system can have more than two independent 
outcomes, and in this case, the measurement cannot be projective. Measurements 
of this general type are essential to an operational approach to quantum 
theory, but so far, the nonprojective character of a measurement could only be 
verified experimentally by already assuming a specific quantum model of parts 
of the experimental setup. Here, we overcome this restriction by using a 
device-independent approach. In an experiment on pairs of 
polarization-entangled photonic qubits we violate by more than 8 standard 
deviations a Bell-like correlation inequality which is valid for all sets of 
two-outcome measurements in any dimension. We combine this with a 
device-independent verification that the system is best described by two 
qubits, which therefore constitutes the first device-independent certification 
of a nonprojective quantum measurement.
\end{abstract}

\maketitle


The qubit is the abstract notion for any system which can be modeled in quantum 
theory by a two-level system. In such a system, any observable has at most two 
eigenvalues and hence any projective measurement can have at most two outcomes. 
Still, a qubit allows for an infinite number of different two-outcome 
measurements, the value of which, in general, cannot be known to the observer 
beforehand, but rather follows a binomial distribution. In quantum information 
theory, additional properties reflecting this binary structure have been 
revealed, e.g., the information capacity of a qubit is one classical bit, even 
when using entangled qubits \cite{Holevo:1973PIT}. Nonetheless, the properties 
of a qubit sometimes break with the binary structure, e.g., transferring the 
quantum state of a qubit is only possible with the communication of two 
classical bits and the help of entanglement \cite{Bennett:1993PRL}. Moreover, 
it is well-known that general quantum measurements can be nonprojective and 
have more than two irreducible outcomes \cite{Busch:1995}. The most general 
quantum measurement with $n$ outcomes is described by positive semidefinite, 
possibly nonprojective, operators $E_1, E_2, \dotsc, E_n$ with $\sum 
E_k=\openone$. The number of outcomes is reducible, if it is possible to write 
$E_k= \sum_\lambda p_\lambda E^{(\lambda)}_k$ so that $E^{(\lambda)}_1, \dotsc, 
E^{(\lambda)}_n$ are measurements for each $\lambda$, $p_\lambda$ is a 
probability distribution over $\lambda$, and for each $\lambda$ there is at 
least one $k_\lambda$ with $E^{(\lambda)}_{k_\lambda}= 0$. Nonprojective 
measurements found first applications in quantum information processing in the 
context of the discrimination of nonorthogonal quantum states. Ivanovic 
\cite{Ivanovic:1987PLA} found that it is possible to discriminate two pure 
qubit states without error even if the two states are nonorthogonal, but at the 
cost of allowing a third measurement outcome that indicates a failure of the 
discrimination procedure. The strategy with the lowest failure probability can 
be shown to be an irreducible three-outcome measurement \cite{Peres:1988PLA}.  
Also recently, nonprojective measurements proved to be essential in purely 
information theoretical tasks like improving randomness certification 
\cite{Acin:2016PRA}.

A peculiarity of nonprojective qubit measurements with more than two 
irreducible outcomes is that there is no known way to implement them within a 
qubit system. Rather, the measurement apparatus needs to manifestly work 
outside of what would be modeled by a qubit alone. To some extent it is 
therefore a matter of perspective whether, at all, one is willing to admit such nonprojective measurements on a qubit system. However, 
device-independent self-testing \cite{Bardyn:2009PRA} allows us to demonstrate 
that a qubit description is appropriate for the tested system, by showing that, 
with high precision, any measurement on the system can be modeled as a qubit 
measurement.

A key observation is that it is not possible to show that a measurement is 
irreducibly nonbinary insofar we consider a single quantum system, as the 
outcomes of measurements on a single system can always be explained in terms of 
a hidden variable model where all $E^{(\lambda)}_k$ are either $\openone$ or 
$0$ and $p_\lambda$ depends on the preparation of the system. The situation 
changes when considering the correlations between independent measurements on 
an entangled system \cite{Bell:1964PHY}, but still, a violation of a 
conventional Bell inequality on qubits---however high---can always be explained 
by locally selecting from binary quantum measurements \cite{Kleinmann:2016PRL}. 
Yet, there are specialized Bell-like inequalities, where qubit measurements 
with more than three outcomes outperform the maximal violation attainable when 
only binary measurements are considered \cite{Vertesi:2010PRA}. An analysis of 
this advantage reveals that this effect is very small and would require an 
overall visibility of more than $0.992$ \cite{Vertesi:2010PRA, Barra:2012PRA, 
Kleinmann:2016PRL}.


\begin{figure}
\includegraphics[width=.95\linewidth]{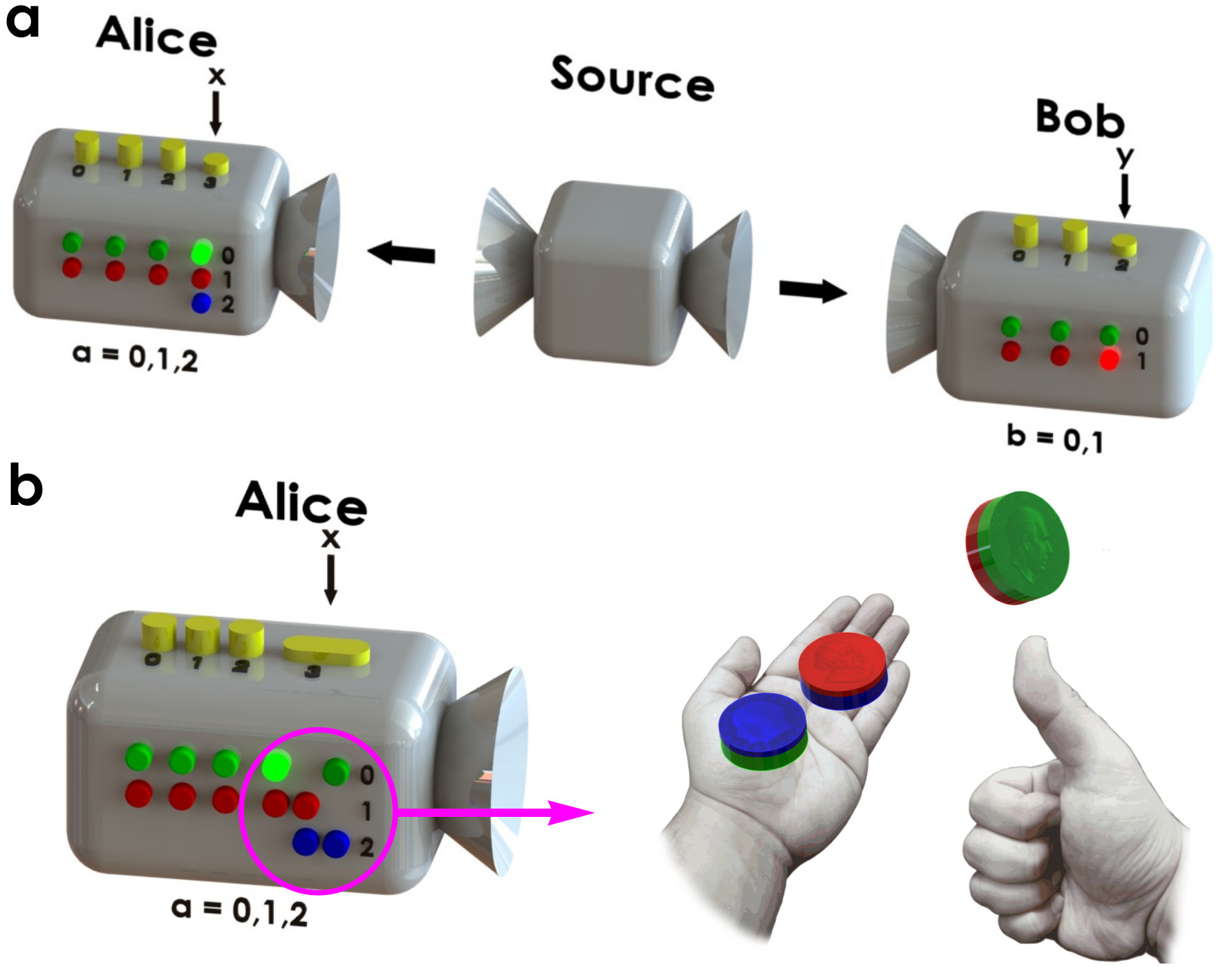}
\caption{\label{Fig1}%
Testing correlations that cannot be explained in terms of binary measurements.
(a) Scheme of the test performed. Pairs of entangled systems are sent to 
Alice's and Bob's laboratories (represented by boxes with yellow buttons at the 
top and lights of different color in the side). In each laboratory one system 
is submitted to a measurement (represented by the yellow button pressed) and 
produces an outcome (represented by a light flashing). All possible 
measurements have two outcomes, except for Alice's measurement $x=3$ which has 
three outcomes (represented by lights of different color, green for 0, red for 
1, and blue for 2).
(b) Discarded scenario. Our experiment excludes that the outcomes of Alice's 
measurement $x=3$ are produced by a measurement apparatus that selects one out 
of three binary quantum measurements with outcomes 0/1, 1/2, or 2/0 
(represented by three coins with green/red, red/blue, and blue/green 
sides, respectively).}
\end{figure}


Here we introduce an inequality where this threshold is lowered to $0.9845$, 
enabling the device-independent certification of a nonbinary measurement on a 
qubit.  We consider a bipartite scenario, cf.\ Fig.~\ref{Fig1}(a), where one 
party, Alice, chooses one among four measurements $x=0,1,2,3$ while the other 
party, Bob, chooses one among three measurements $y=0,1,2$. All measurements 
have two outcomes, $a=0,1$ and $b=0,1$, except Alice's measurement $x=3$, which 
has three outcomes, $a=0,1,2$. We denote by $P(ab|xy)$ the probability for 
outcome $a$ and $b$ when the setting $x$ and $y$ were chosen and consider the 
expression
\begin{multline}\label{Ineq}
I= P(00|00)+P(00|11)+P(00|22)\\ -P(00|01)-P(00|12)-P(00|20)\\ 
-P(00|30)-P(10|31)-P(20|32).
\end{multline}
When restricted to binary quantum measurements, not necessarily on a qubit, 
then the value of $I$ is upper bounded by $1.2711$. Without this restriction, the maximal quantum value of 
$I$ is $3\sqrt3/4\approx1.2990$ and can be achieved for two qubits using a 
maximally entangled state. Thus, an 
experiment violating the inequality $I<1.2711$ proves that Alice's measurement 
$x=3$ cannot have been a measurement composed of binary quantum measurements on 
whatever quantum system and selected by the measurement apparatus, as shown in 
Fig.~\ref{Fig1}(b).

Since projective measurements on a qubit necessarily are binary or trivial, a 
violation of $I<1.2711$ certifies the implementation of a nonprojective 
measurement. This requires, however, that the system at Alice's laboratory is 
actually a qubit, which is manifestly the case in our experimental set-up, as 
we explain below. In addition, this assertion of Alice's system being a qubit, 
can also be verified in a device-independent way by measuring the violation of 
the Clauser–Horne–Shimony–Holt (CHSH) Bell inequality \cite{Clauser:1969PRL}.  
If this violation is maximal, the joint state has to be a maximally entangled 
qubit-qubit state \cite{Summers:1987CMP, Popescu:1992PLA, Tsirelson:1993HJS}, 
independently of what measurement apparatuses are used. If the observed value 
for the CHSH violation deviates by $\epsilon$ from the maximum $2\sqrt2-2$, the 
state must still have a fidelity of at least $1-2.2\epsilon$ with a maximally 
entangled qubit-qubit state \cite{Bancal:2015PRA}. A description of the system 
in the corresponding qubit-qubit-space is hence accurate up to $2.2\epsilon$.


\begin{figure*}
\includegraphics[width=.95\linewidth]{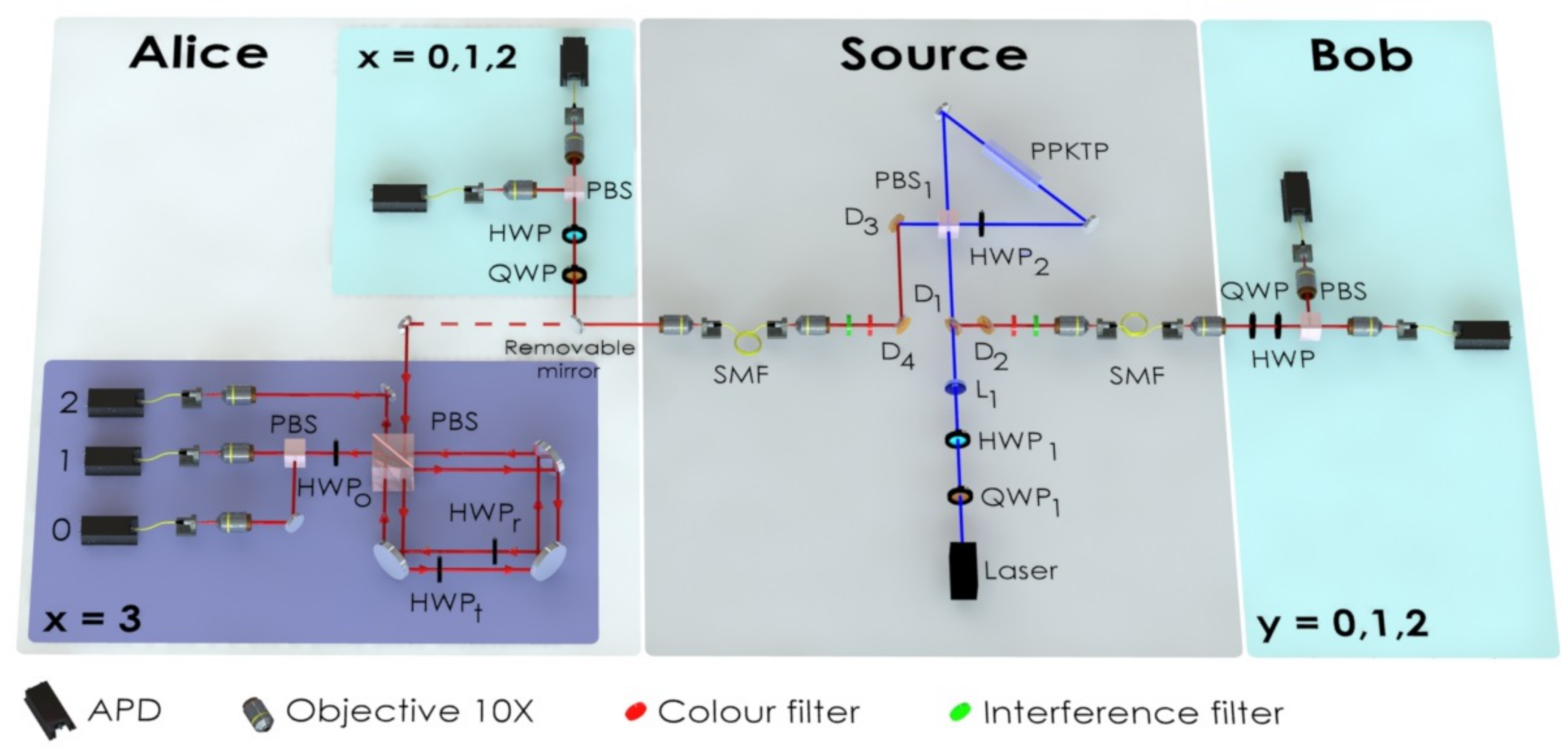}
\caption{\label{Fig2}%
Experimental set-up.
A PPKTP nonlinear crystal placed into a phase-stable Sagnac interferometer is 
pumped by a single mode laser operating at \unit[405]{nm} to produce pairs of 
polarization-entangled photons at \unit[810]{nm}. The quarter-wave plate 
QWP$_1$ and the half-wave plate HWP$_1$ are used to control the polarization 
mode of the pump beam. Dichroic mirrors (D) and longpass color filters are used 
to remove the pump beam light. The generated photons are then sent to Alice and 
Bob through single-mode fibers (SMF). Alice (Bob) can choose among three 
different binary measurements (depicted in blue boxes) labeled by $x=0,1,2$ 
($y=0,1,2$). These measurements are performed using a set of a QWP, a HWP, and 
a PBS. Besides, Alice also performs a three-outcome measurement $x=3$ using a 
polarization based two-path Sagnac interferometer (depicted in the Alice's 
violet box). The elements of the three-outcome qubit measurement are defined by 
HWP$_r$, HWP$_t$, and HWP$_o$. The coincidence counts between Alice's and Bob's 
detectors are recorded using a coincidence electronics unit based on a field 
programmable gate array device.}
\end{figure*}


The set-up of our experiment is shown in Fig.~\ref{Fig2}. Degenerate 
\unit[810]{nm} photon pairs, with orthogonal polarisations, are produced from 
spontaneous parametric down-conversion (SPDC) in a bulk type-II nonlinear 
periodically poled potassium titanyl phosphate (PPKTP) \unit[20]{mm} long 
crystal. The crystal is pumped by a single-longitudinal mode continuous wave 
\unit[405]{nm} laser with \unit[1]{mW} of optical power. We resort to an 
ultra-bright source architecture, where the type-II nonlinear crystal is placed 
inside an intrinsically phase-stable Sagnac interferometer \cite{Kim:2006PRA, 
Wong:2006LP, Fedrizzi:2007OE}. This interferometer is composed of two laser 
mirrors, a half-wave plate (HWP$_2$) and a polarizing beamsplitter cube 
(PBS$_1$). HWP$_2$ and PBS$_1$ are both dual-wavelength with anti-reflection 
coatings at \unit[405]{nm} and \unit[810]{nm}. The fast axis of the HWP$_2$ is 
set at \unit[45]{degree} with respect to the horizontal, such that 
down-converted photons are generated in the clockwise and counter-clockwise 
directions. The clockwise and counter-clockwise propagating modes overlap 
inside the polarizing beamsplitter and, by properly adjusting the pump beam 
polarization mode, the two-photon state emerging at the output ports is 
$\ket{\psi^+}= (\ket{HV}+\ket{VH}) /\sqrt2$, where $\ket{H}$ ($\ket{V}$) 
denotes the horizontal (vertical) polarization of a down-converted photon. Due 
to the phase-matching conditions, there may be entanglement between other 
degrees of freedom of the generated photons, or coupling between the 
polarization and the momentum of these photons that would compromise the 
quality of the polarization entanglement. To avoid this we add extra spectral 
and spatial filtering. To remove the remaining laser light we adopt a series of 
dichroic mirrors followed by a longpass color glass filter. Then, Semrock 
high-quality (peak transmission greater than \unit{90}\%) narrow bandpass 
(full-width-half-maximum of \unit[0.5]{nm}) interference filters centered at 
\unit[810]{nm} are used to ensure that phase-matching conditions are achieved 
with the horizontal and vertical polarization modes at degenerated frequencies.

The indistinguishability of the photon pair modes (``HV'' and ``VH'') is 
guaranteed by coupling the generated down-converted photons into single mode 
fibers. These fibers implement a spatial mode filtering of the down-converted 
light, destroying any residual spatial entanglement or polarization-momentum 
coupling. To maximize the source's spectral brightness, we resort to a 
numerical model \cite{Ljunggren:2005PRA}. In our case, the beam waist $w_p$ of 
the pump beam, and $w_\mathrm{SPDC}$ of the selected down-converted modes, at 
the center of the PPKTP crystal, are adjusted by using a \unit[20]{cm} focal 
length lens (L$_1$) and 10$X$ objective lenses. The optimal condition for 
maximal photon-par yield is obtained when $w_\mathrm{SPDC}=\sqrt2 w_p$, with 
$w_p=\unit[40]{\mu m}$. The observed source spectral brightness was $410000$ 
photon pairs \unit{(s mW nm)$^{-1}$}. The quality of the polarization 
entanglement generated at the source site was measured by observing a mean 
two-photon visibility of $0.987\pm0.002$ while measuring over the logical and 
diagonal polarization bases.

Due to the demand of a high overall visibility we built a coincidence 
electronics based on a field programmable gate array platform and capable of 
implementing up to \unit[1]{ns} coincidence windows, thus reducing the 
probability for an accidental coincidence count to less than $0.00025$. 
Therefore, the evaluation of the data does not require a separate treatment for 
accidental coincidence counts. The down-converted photons are registered using 
PerkinElmer single-photon avalanche detectors with an overall detection 
efficiency of \unit{15}\%. We account for this by including the assumption into 
our analysis that the detected coincides are a fair sample from the set of all 
photon pairs.

Alice's and Bob's binary measurements are implemented using a set composed of a 
quarter wave plate (QWP), a HWP, and a PBS for each party, cf.\ 
Fig.~\ref{Fig2}. A high-quality film polarizer is also used in front of the 
detectors (not shown for sake of clarity) to obtain a total extinction ratio of 
the polarizers equal to $10^7$:$1$. Therefore, in our experiment the two-photon 
visibility is not upper limited by the polarization contrast of our measurement 
apparatuses. Alice's three-outcome measurement $x=3$ is implemented using the 
propagation modes of Alice's down-converted photon. With this purpose, Alice's 
photons are sent, after displacing a removable mirror, through a polarization 
based two-path Sagnac interferometer. The propagation modes of a photon within 
this interferometer are not co-propagating and depend on its polarization 
state. This allows for conditional polarization transformations implemented 
with HWPs placed inside the interferometer, as shown in Fig.~\ref{Fig2}. The 
plate located at the clockwise reflected mode is denoted by HWP$_r$ and the 
plate at the counter-clockwise transmitted mode by HWP$_t$. The fast axis of 
HWP$_r$ is oriented in the direction of the horizontal axis, while HWP$_t$ is 
oriented at an angle of \unit[117.37]\degree. The two propagation modes are 
then superposed again at the PBS, and at one of the output ports of the 
interferometer an extra HWP$_o$, oriented at \unit[112.5]\degree, and a PBS is 
used to conclude the three-outcome measurement. Further details on the 
implementation of the Alice's three-outcome measurement $x=3$ are given in the 
appendix.


\begin{figure}
\includegraphics[width=.95\linewidth]{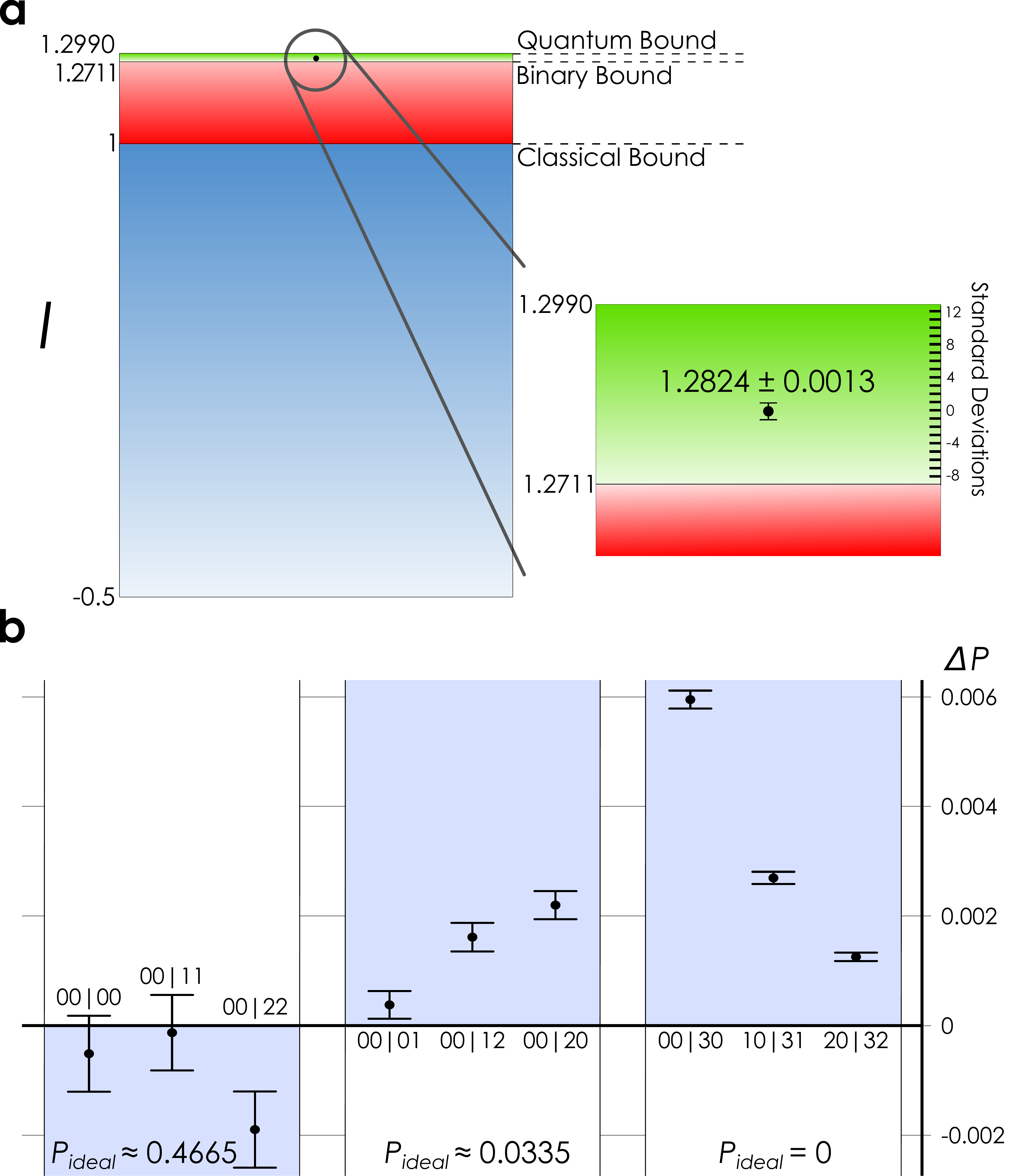}
\caption{\label{Fig3}%
Experimental results.
(a) Critical values and experimental value of $I$, cf.\ Eq.~\eqref{Ineq}. 
Uncorrelated, uniformly random events yield $-1/2$, general local hidden 
variable models cannot exceed 1, binary quantum measurements do not reach 
$1.2711$, and $3\sqrt{3}/4\approx 1.2990$ is the universal bound according to 
quantum theory. The experimental value obtained is $I=1.2824\pm0.0013$, 
violating the bound for binary quantum measurements by more than 8 standard 
deviations.
(b) Correlation measurements. For each correlation $P(ab|xy)$ in $I$, the 
deviation of the measured value from the ideal value is shown, $\Delta 
P=P_\mathrm{exp}-P_\mathrm{ideal}$. Deviations in the blue shaded areas 
decrease the experimental value of $I$. Error bars correspond to 1 standard 
deviation and are calculated assuming fair samples from Poissonian 
distributions.}
\end{figure}


In the experiment, the correlations $P(ab|xy)$ in $I$ were measured by 
integrating coincidences over a time of \unit[240]s for each outcome and 
normalizing over the total number of coincidences per setting. The results are 
shown in Fig.~\ref{Fig3} and yield a measured value of $I=1.2824\pm0.0013$. The 
measurement settings were implemented independently for Alice and Bob, 
justifying the assumption that Alice's measurements also act independently of 
Bob's measurement setting $y$ and vice versa. Hence, any explanation in terms 
of binary measurements on an arbitrary quantum system is excluded by $8.7$ 
standard deviations, which corresponds to a p-value of $1.6\times 10^{-18}$.

In order to prove that Alice's measurement $x=3$ is a nonprojective 
quantum measurement, we need also to verify that Alice's system can be properly 
described as a qubit. We rely here on two complementary arguments. First, one 
can resort to the design of the experiment where the source is designed to 
produce entanglement in polarization, i.e., qubit-qubit entanglement. Second, 
we measured the CHSH correlations with our set-up and observed a violation of 
$2\sqrt2-2-\epsilon$ with $\epsilon=0.0253\pm0.0014$ and hence the fidelity 
with a maximally entangled qubit-qubit state is guaranteed in a 
device-independent way to be at least $0.9351$ within $3$ standard deviations 
\cite{Bancal:2015PRA}. Note, that we measured the CHSH correlations using the 
same source and the same measurement set-up as we used for the measurement of 
$I$---except that different angles at the HWPs are adopted. Except for some 
ubiquitous adversary ad-hoc models we can hence conclude that also in the 
measurement of $I$, the fidelity of the state with a maximally entangled 
qubit-qubit state is at least $0.9351$. Notice that the estimate for the 
fidelity is pessimistic since imperfections in the measurement apparatuses 
reduce the CHSH violation and therefore lower the bound on the fidelity. Still, 
in an alternative explanation where \unit{93.51}\% of the times binary 
measurement was used, a bound of $I< 0.9351\times1.2711+0.0649\times 3\sqrt3/4 
< 1.2730$ would have to be obeyed, which is clearly violated in the experiment.

Our result shows that three-outcome nonprojective measurements can produce 
strictly stronger correlations between two qubits than projective two-outcome 
measurements on any quantum system and, therefore, that nature cannot be 
described in terms of binary quantum tests, not even when these tests are 
performed on two-level quantum systems. Quantum theory predicts also 
qubit-qubit correlations that cannot be explained as produced by three-outcome 
measurements. Observing them requires an overall visibility above $0.9927$, 
which is beyond what is currently feasible in our set-up.  Further theoretical and experimental efforts will be 
needed to identify and produce qubit correlations which can only be explained 
by four-outcome nonprojective measurements. This will be the farthest we can 
go, as qubit correlations can always be accounted that way 
\cite{Chiribella:2007PRL}.

\begin{acknowledgements}
We thank Antonio Acín,
Marcelo Terra-Cunha,
Paolo Mataloni,
Valerio Scarani,
Jaime Cariñe,
Miguel Solís-Prosser and
Omar Jiménez for conversations and assistance.
This work was supported by
FONDECYT grants 1160400, 11150325, 11150324, 1140635, 1150101, Milenio grant 
RC130001, PIA-CONICYT grant PFB0824, OTKA grant K111734, the FQXi large grant 
project ``The Nature of Information in Sequential Quantum Measurements'', the 
project FIS2014-60843-P, ``Advanced Quantum Information'' (MINECO, Spain), with 
FEDER funds, the Knut and Alice Wallenberg Foundation, Sweden (Project 
``Photonic Quantum Information''), the EU (ERC Starting Grant GEDENTQOPT), and 
the DFG (Forschungsstipendium KL 2726/2-1). P.G.\ and J.F.B.\ acknowledge the 
financial support of CONICYT.
\end{acknowledgements}



\appendix
\section{Maximal value of $I$ for binary quantum measurements.}
\label{appA}

To obtain the bound for $I$ while considering only binary quantum measurements,
we note that $I$ contains the chained Bell inequality \cite{Pearle:1970PRD,
Braunstein:1990AP} $I_\mathrm{chain}\le1$ with three settings, where
\begin{multline}
I_\mathrm{chain}= P(00|00)+P(00|11)+P(00|22)\\ -P(00|01)-P(00|12)-P(00|20).
\end{multline}
The remainder, $I-I_\mathrm{chain}= -P(00|30)-P(10|31)-P(20|32)$ only involves
correlations of Alice's three-outcome measurement $x=3$. There are three
possibilities for replacing Alice's measurement $x=3$ by a binary measurement,
by omitting $a=0$, $a=1$, or $a=2$. Taking into account the permutation
symmetry of $I$, all of them are equivalent to $I'=
I_\mathrm{chain}-P(00|30)-P(10|31)$. We used the Navascués–Pironio–Acín (NPA)
hierarchy \cite{Navascues:2007PRL} to obtain an upper bound on the maximal
value $I'$. Running level 2 of the hierarchy, we obtain 1.271045 for this
bound. Within the numerical precision, this value can be attained with a
partially entangled qubit-qubit state showing that 1.2711 also corresponds to
the maximal value of $I$ with binary qubit measurements.


\section{Maximal value of $I$ for arbitrary quantum measurements.}
\label{appB}

An upper bound on the maximal value of $I$ attainable in quantum theory can be
obtained by upper bounding $I_\mathrm{chain}$ and the remainder
$I-I_\mathrm{chain}$ separately. The maximal value of $I_\mathrm{chain}$ is
$3\sqrt3/4$ and can be attained with a qubit-qubit maximally entangled state
\cite{Wehner:2006PRA}. On the other hand, by construction, $I-I_\mathrm{chain}$
cannot be greater than zero since it only contains nonpositive terms. Put
together, the maximal value of $I$ is upper bounded by $3\sqrt3/4$.

This value is tight and can be attained by preparing the qubit-qubit state
$\ket{\psi^+}= (\ket{01}+\ket{10})/\sqrt2$ and choosing the following
measurements: Alice's binary measurements $x=0,1,2$ are defined by $M_{0|x}=
P(\alpha_x)$ and $M_{1|x}= \openone-P(\alpha_x)$, with $P(\theta)= (\openone+
\sigma_z\cos\theta+ \sigma_x\sin\theta)/2$, where $\sigma_z$ and $\sigma_x$ are
Pauli matrices, and the angles are given by $\alpha_0=3\pi/2$,
$\alpha_1=\pi/6$, and $\alpha_2=5\pi/6$. Alice's three-outcome measurement
$x=3$ is defined by $M_{a|3}= 2P(\gamma_a)/3$ for $a=0,1,2$, with angles
$\gamma_0=2\pi/3$, $\gamma_1=4\pi/3$, and $\gamma_2=0$. Bob's measurements are
defined by $M_{0|y}= P(-\gamma_y)$ and $M_{1|y}= \openone-P(-\gamma_y)$ for
$y=0,1,2$.


\section{Implementation of Alice's three-outcome measurement.}
\label{appC}

The three-outcome measurement is implemented by sending Alice's down-converted
photon through a polarization based two-path Sagnac interferometer, cf.\
Fig.~\ref{Fig2}. We write $\ket H$ ($\ket V$) for the horizontal (vertical)
polarization. The mode entering the interferometer, rotating counter-clockwise
and leaving for outcomes 0 and 1 is denoted by $\ket a$. $\ket b$ denotes the
clockwise rotating mode leaving for outcome 3. In this way, the action of the
PBS within the interferometer is given by
\begin{multline}
U_\mathrm{PBS}= \ketbra HH (\ketbra aa+\ketbra bb)\\ +i\ketbra VV (\ketbra
ab+\ketbra ba).
\end{multline}
The actions of the HWP$_t$ and HWP$_r$ of the interferometer in the transmitted
and reflected mode, respectively, combine to $U_{t,r} =
U_\mathrm{HWP}(\gamma'_t) \ketbra aa + U_\mathrm{HWP}(\gamma'_r) \ketbra bb$,
where $U_\mathrm{HWP}(\gamma')$ is the Jones matrix of a HWP whose fast axis is
oriented at an angle $\gamma'$ with respect to the horizontal axis
\begin{multline}
U_\mathrm{HWP}(\gamma')= \cos(2\gamma')(\ketbra HH-\ketbra VV)\\
+\sin(2\gamma')(\ketbra VH+\ketbra HV).
\end{multline}
Therefore, the Sagnac interferometer is described by $U_\mathrm{S}=
U_\mathrm{PBS} U_{t,r} U_\mathrm{PBS}$.

After the interferometer, the photon in mode $\ket a$ is transmitted through
HWP$_o$ and an additional PBS. On the polarization degree of freedom, the three
outcome modes 0, 1, and 2 are hence mediated by $\ket\psi \rightarrow
A_k\ket\psi$ with the Kraus operators
\begin{equation}\begin{split}
A_0&= \braket{b|U_\mathrm{PBS}U_\mathrm{HWP}(\gamma'_o)\ketbra aa
U_\mathrm{S}|a},\\
A_1&= \braket{a|U_\mathrm{PBS}U_\mathrm{HWP}(\gamma'_o)\ketbra aa
U_\mathrm{S}|a}\text{, and }\\
A_2&= \braket{b|U_\mathrm{S}|a},
\end{split}\end{equation}
so that the implemented three-outcome measurement is given by $M_{k|3}=A_k^\dag
A_k$.  The measurement required for a maximal violation of $I$ is achieved with
$\gamma'_r=0$, $\gamma'_t\approx 117.37\degree$, and $\gamma'_o=112.5\degree$.

\section{Qubit-qubit correlation inexplicable by three-outcome nonprojective 
measurements.}
\label{appD}
We consider a scenario where Alice chooses among the binary measurements
$x=0,1,2$ and the four-outcome measurement $x=3$ and Bob chooses among the
binary measurement $y=0,1,2,3$. The expression
\begin{equation}
L=\beta_\mathrm{el}-8\sum_{i=0}^3P(i,0|3,i),
\end{equation}
has been used in Ref.~\cite{Acin:2016PRA} in the context of randomness
extraction.
The term $\beta_\mathrm{el}$ was introduced by Bechmann-Pasquinucci and Gisin
\cite{Bechmann:2003PRA} in the Bell inequality $\beta_\mathrm{el}\le6$, where
\begin{multline}
\beta_\mathrm{el}=
+ P(10|02)
+ P(10|03)
+ P(10|11)
+ P(10|13)\\
+ P(10|21)
+ P(10|22)
+2P(00|00)\\
+2P(00|10)
+2P(00|20)
+4P(00|01)\\
+4P(00|12)
+4P(00|23)
-2P(10|00)\\
-2P(10|10)
-2P(10|20)
-3P(00|02)\\
-3P(00|03)
-3P(00|11)
-3P(00|13)\\
-3P(00|21)
-3P(00|22).
\end{multline}
Applying the methods developed in Section~\ref{appA} and Section~\ref{appB},
one finds, using the third level of the NPA-hierarchy, that the value of $L$ is
upper bounded by 6.6876 for binary measurement and by 6.8489 for three-outcome
measurements. Using four-outcome qubit measurements, $L$ can reach a value of
$4\sqrt 3>6.9282$. Therefore, a verification of an irreducible four-outcome
qubit measurements requires a visibility of 0.9928.

\bibliography{article}

\end{document}